\documentclass[aps,11pt,superscriptaddress,nofootinbib]{revtex4}

\usepackage{graphicx}
\usepackage{amsmath}
\usepackage{mathrsfs}
\usepackage{bm}
\usepackage{verbatim}

\newcommand{\beq}{\begin{equation}}
\newcommand{\eeq}{\end{equation}}
\newcommand{\bea}{\begin{eqnarray}}
\newcommand{\eea}{\end{eqnarray}}

\newcommand{\bvec}[1]{{\bm #1}} 

\newcommand{\simge}{\hspace*{0.2em}\raisebox{0.5ex}{$>$}
     \hspace{-0.8em}\raisebox{-0.3em}{$\sim$}\hspace*{0.2em}}
\def\simle{\hspace*{0.2em}\raisebox{0.5ex}{$<$}
     \hspace{-0.8em}\raisebox{-0.3em}{$\sim$}\hspace*{0.2em}}

\begin{document}

\title{Renormalization of Singular Potentials and Power Counting}

\author{B. Long}
\affiliation{Department of Physics, University of Arizona, 
Tucson, AZ 85721, USA \vspace{0.2cm}}
\author{U. van Kolck}
\affiliation{Department of Physics, University of Arizona, 
Tucson, AZ 85721, USA \vspace{0.2cm}}
\affiliation{Kernfysisch Versneller Instituut,
        Rijksuniversiteit Groningen,
	Zernikelaan 25, 9747 AA Groningen, The Netherlands
        \vspace{0.2cm}}
\affiliation{Instituto de F\'\i sica Te\'orica,
        Universidade Estadual Paulista,
	Rua Pamplona 145, 01405-900 S\~ao Paulo, SP, Brazil
	\vspace{0.2cm}}
\date{\today}

\begin{abstract}
We use a toy model to illustrate how to build 
effective theories for singular potentials. 
We consider a central attractive $1/r^2$ potential  
perturbed by a $1/r^4$ correction.
The power-counting rule, an important ingredient of effective theory,
is established by seeking the minimum set of short-range
counterterms that renormalize the scattering amplitude. 
We show that leading-order counterterms are 
needed in all partial waves where the potential overcomes 
the centrifugal barrier, and that the additional counterterms 
at next-to-leading order
are the ones expected on the basis of dimensional analysis.
\end{abstract}


\maketitle

\section{\label{sec:intro} Introduction}
Singular potentials ---those that diverge as $1/r^n$, $n\ge 2$ at
a small distance $r$--- are common in atomic, molecular
and nuclear physics \cite{frank_land_spector}. 
Contrary to the regular case, a sufficiently attractive singular potential 
does not determine observables uniquely \cite{case}. 
A proper formulation of the extra ingredients can be found
in the framework of effective
field theory (EFT) \cite{beane}, where 
the problem is cast in the usual language of renormalization. 
With the techniques of renormalization one retains the
predictive power of singular potentials at low energies, 
while short-distance physics is accounted for in
a minimal, model-independent way.

The basic idea is
that in order to obtain the potential from
some underlying theory we make an arbitrary decomposition
between short- and long-distance physics through  a short-distance
cutoff $R\sim 1/\Lambda$, above which the potential
has a form determined by the exchange of light particles
---such as two-photon exchange in the case of van der Waals forces,
or pion exchange in nuclear physics.
When the long-range potential is regular, or singular and repulsive,
the quantum-mechanical dynamics is insensitive to $R$.
When the potential is singular and attractive, on the other hand,
observables would
depend sensitively on $R$ 
if we ignored short-distance physics.
Dependence on the arbitrary scale $R$ can be removed  \cite{beane} by 
adjusting parameters of the potential at short distances
as functions of $R$ (``running constants'')
to guarantee that low-energy observables are reproduced
independently of $R$.
Such short-range ``counterterms'' are thus used to mimic the effects
of short-distance physics.

The case of the $S$ wave in an attractive singular central potential
$1/r^n$ 
was considered in Ref. \cite{beane},
and the particular cases $n=2$ and $n=4$ were studied 
in more detail in Refs. \cite{coon,hammer} and \cite{mary}, respectively. 
It was shown that a single
counterterm ---for example, in the form of the depth of a square-well---
is sufficient to ensure an approximate $R$ independence of low-energy
observables, once one observable is used to determine the
running of the counterterm.
An interesting property of this running,
first noticed in the three-body system \cite{3bozos},
is an oscillatory behavior characteristic of limit cycles. 
For example, when $n=2$ the short-distance force has a fixed period
as a function of $\ln R$.
Limit cycles are now a subject of renewed interest
\cite{limits}.

A consistent EFT must in addition provide systematic improvement 
over this simple picture.
If we consider particles of typical momentum $Q$,
we would like to be able to calculate
observables in an expansion in powers of $Q R_{sh}= Q/M_{hi}$,
where $R_{sh}\equiv 1/M_{hi}$ is the range of short-range physics.
Successive terms in this expansion are referred to as
leading order (LO), next-to-leading order (NLO), and so on.
Determining at which orders interactions contribute
is called ``power counting''.

We are interested here in a two-body system in a regime of momenta where
LO comes from a dominant long-range potential that is attractive
and singular, 
together with the required counterterms.
The latter include one $S$-wave counterterm \cite{beane}.
However, the long-range potential contributes also to other partial waves,
and it has been argued on general terms \cite{behncke68} that 
counterterms are required in those waves as well.
In the particular case of nucleons,
one-pion exchange has an
$1/r^3$ tensor force in spin-triplet channels;
it has been shown explicitly
that counterterms are necessary 
in all waves where the tensor force is attractive and iterated 
to all orders \cite{towards,nogga_timmermans_vankolck,spaniards1}.
(For a different point of view, see Ref. \cite{germans}.)
This poses a potential problem because predictive power seems
to be lost in systems with more than two particles,
where all two-particle waves contribute.
In Ref. \cite{nogga_timmermans_vankolck} a solution
was proposed where, thanks to the centrifugal barrier, perturbation theory 
in the long-range potential is employed in waves
of sufficiently large angular momentum.
We would like to better understand the role of angular momentum, and 
establish the number of short-range counterterms needed in LO to make sense
of a singular potential. 

In general a dominant long-range potential suffers
corrections that increase as the distance decreases.
These corrections can arise from additional, smaller couplings
to the light exchanged particles ---e.g. magnetic photon couplings
of the Pauli type.
They can also be generated by the simultaneous exchange of several light
particles ---such as two-pion exchange. 
These effects lead to
potentials that fall faster at
large distances than the LO potential.
For example, an LO $1/r^n$ potential
might have an NLO correction $R_{sh}^2/r^{n+2}$, as
is the case for
two-pion exchange between nucleons, 
which (neglecting nucleon excitations) goes \cite{weinberg,others,sameold}
as  $R_{sh}^2/r^{5}$ 
with $R_{sh}=(4\pi f_\pi)^{-1}\simeq 0.2$ fm
the characteristic QCD scale. 
These potentials, which are subleading at distances $r\simge R_{sh}$,
are more singular than the LO potential and overcome it
for $r\simle R_{sh}$.
The enhanced singularity ought to demand new counterterms.
The desired expansion in $Q R_{sh}$ requires that
NLO counterterms, which represent physics at $r\simle R_{sh}$,
balance the NLO long-range potential ---that is, the full NLO
should be such that changes in observables are small.
If that is the case, we expect  \cite{towards} that one can treat the NLO
potential in perturbation theory.
One would also expect \cite {nogga_timmermans_vankolck} 
the number of counterterms to be given
by naive dimensional analysis ---for example, that
an NLO correction $R_{sh}^2/r^{n+2}$ needs counterterms 
with two more derivatives than the counterterms required by 
the LO $1/r^n$.
In the nuclear case, existing calculations
have followed \cite{sameold}
instead the original suggestion \cite{weinberg}
that subleading corrections in the potential be iterated to all orders.
In fact, it has been argued \cite{spaniards2} that
this requires fewer counterterms than doing perturbation theory
on the corrections.
Nevertheless, treating small corrections in perturbation
theory is conceptually simpler, as
the running of the LO counterterms is not completely modified by the
higher singularity of the NLO potential.
Understanding the strengths and limitations of a perturbative approach would 
at 
the very least help delineate the scope of a resummation of NLO corrections.

In this paper we address the issue of power counting for singular potentials,
in particular the role of centrifugal forces in LO
and the perturbative renormalizability of NLO interactions.
We examine this issue in a simple toy model,
where the LO and NLO long-range potentials are taken as (central)
$-1/r^2$ and $\pm 1/r^4$, respectively.
Some of our arguments are similar to those
employed in related two \cite{coverass2} and three \cite{coverass3,
  nnlodispute, birse} -body contexts.
We show that 
LO counterterms are required in all partial
waves up to a critical value,
and that the number of NLO counterterms is just what
is expected on the basis of dimensional analysis.
Most of our conclusions can be extended to other
attractive singular potentials.

This paper is organized as follows. In Sect. \ref{sec:frame}
we discuss our EFT framework.
In Sect. \ref{sec:LO} a new
approach for treating the $-1/r^2$ potential is presented. 
In Sect. \ref{sec:NLO} the renormalization of the NLO amplitude is discussed
and, as a result, NLO counterterms are found. 
We discuss some of the implications
of our results to other potentials in Sect. \ref{sec:discussion}.
Finally, we summarize our findings
in Sect. \ref{sec:summary}.

\section{\label{sec:frame} Framework}
We consider two non-relativistic particles of reduced mass $m$ 
in the center-of-mass frame,
which interact through a potential $V$.
Our analysis will mainly be based upon the Lippmann-Schwinger equation
of the half-off-shell $T$ matrix,
\beq
T(\bvec{p}',\bvec{p}) =  V(\bvec{p}', \bvec{p})
         + 2m \int \frac{d^3 q}{(2\pi)^3} 
                   \frac{V(\bvec{p}', \bvec{q})}{p^2- q^2 + i\epsilon}
                   T(\bvec{q}, \bvec{p}) \,,
\label{eqn:LS0}
\eeq
where $\bvec{p}$ ($\bvec{p}'$) is the initial (final) -state relative momentum.
The physics of angular momentum is most transparent when we use a partial-wave
decomposition. Our convention is that a quantity $O$ is given
in terms of its partial-wave projection $O_l$ by
\beq 
O(\bvec{p}', \bvec{p})= \sum_{l=0}^{\infty} (2l+1) O_l(p', p) P_l(\cos\theta ) 
\, ,
\eeq
where $\theta$ is the angle between $\bvec{p}$ and $\bvec{p}'$.
The partial-wave version of Eq. (\ref{eqn:LS0}) is
\bea
T_l(p', p) &=& V_l(p', p) 
   -\frac{m}{\pi^2} \int_0^{\Lambda} dq \frac{q^2}{q^2 - p^2- i\epsilon}  
   V_l(p',q) T_l(q, p) \, .
\label{eqn:LSlnocc}
\eea
Here we have inserted the ultraviolet cutoff
$\Lambda$, which is in general needed to obtain a well-defined solution.
Since $\Lambda$ is arbitrary, observables (obtained from the
on-shell $T$ matrix)
should be independent of
$\Lambda$, 
\beq
\Lambda \frac{d}{d \Lambda} T\left(p, p\right)=0 \; ,
\label{eqn:rge}
\eeq
that is, renormalization-group (RG) invariant.
It proves convenient to introduce the reduced partial amplitude
\beq
t_l(p', p)\equiv \frac{mp}{\pi^2}T_l(p', p)\, ,
\eeq
in terms of which Eq. (\ref{eqn:LSlnocc}) becomes
\bea
t_l(p', p) &=& \frac{mp}{\pi^2}  V_l(p', p)
   -\int_0^{\Lambda} dq \frac{q}{q^2 - p^2- i\epsilon}  
   \frac{mq}{\pi^2} V_l(p',q)  t_l(q, p) \, .
\label{eqn:LSlnoccRED}
\eea
We will also use in numerics the
$K$ matrix because of its reality. The $K$ matrix satisfies the same
integral equation as $T$ but with $i\epsilon$ replaced by the
principal-value prescription. The reduced form
for the partial-wave-projected $K$ matrix,
\beq
k_l(p', p)\equiv \frac{mp}{\pi^2}K_l(p', p)\, ,
\eeq
satisfies
\beq
k_l(p', p) = \frac{mp}{\pi^2}  V_l(p', p) 
   - \mathscr{P} \int_0^{\Lambda} dq
     \frac{q}{q^2 - p^2}\frac{mq}{\pi^2}  V_l(p',q)k_l(q, p)
\, .
\label{eqn:LS-K}
\eeq
The on-shell $k_l(p, p)$ is related to $l$-wave phase shift $\delta_l(p)$ by
\beq
k_l(p, p) \equiv \frac{\tan \delta_l(p)}{\pi^2} \, .
\eeq

We assume a simple central potential,
whose dominant long-range component is singular and attractive.
We write the potential as
\beq
V(\bvec{p}', \bvec{p})= V_L(\bvec{p}', \bvec{p}) + V_S(\bvec{p}', \bvec{p})
\label{eqn:LandS}
\eeq
in terms of a long-range component $V_L$ and a short-range component $V_S$.
The short-range component $V_S(r)$ is a series of 
derivatives of delta functions, so $V_S(\bvec{p}', \bvec{p})$
is a power series in $\bvec{p}^2$, $\bvec{p}'^2$, 
and $\bvec{p}\cdot \bvec{p}'$, starting with a constant.
For definiteness, we take the long-range component to be
\beq
V_L(r) = V_L^{(0)}(r)+ V_L^{(1)}(r)\, .
\label{eqn:VLsplit}
\eeq
Here $V_L^{(0)}(r)$ is an attractive inverse-square potential,
\beq
V_L^{(0)}(r)=-\frac{\lambda}{2 m r^{2} } \, ,
\label{eqn:pot_cordn}
\eeq
with $\lambda > 0$ a dimensionless parameter.
Its Fourier transform is
\beq
V_L^{(0)}(\bvec{p}',\bvec{p})= -\frac{\pi^2\lambda}{m |\bvec{p}'-\bvec{p}|}
\, ,
\label{eqn:pot_mom}
\eeq
and its partial-wave projection,
\beq \label{Vl}
V_{L\, l}^{(0)}(p', p) = - \frac{\pi^2 \lambda}{m\, (2l+1)}\, 
   \left[\theta(p'-p)\, \frac{p^l}{p'^{l+1}} +
     \theta(p-p')\,\frac{p'^l}{p^{l+1}} \right]
 = - \frac{\pi^2 \lambda}{m\, (2l+1)}\, \frac{p_<^l}{p_>^{l+1}}
\, ,
\eeq
where $p_> \equiv \max\{ p', p \}$ and $p_< \equiv \min\{ p', p\}$. 
In addition,
$V_L^{(1)}(r)$ is an inverse-quartic potential,
\beq
V_L^{(1)}(r)=-\frac{g}{2m M^2 r^4} \, ,
\label{eqn:coor_r4}
\eeq
with $M$ a mass scale and $g$ another dimensionless parameter.
$V_L^{(1)}$ is more singular than $V_L^{(0)}$,
which in momentum space is reflected in higher powers of momenta
in the numerator. To define the Fourier transform we 
need to limit the integration to distances larger than a
coordinate-space cutoff $R$:
\bea
V_{L\, l}^{(1)}(p', p;R) &=& 4\pi \int_R^\infty dr r^2 
                             j_l(p'r) V^{(1)}(r) j_l(pr) 
\nonumber\\
&=& - \frac{\pi^2 g}{2m M^2}\left\{
 \frac{4}{\pi R} \delta_{l0}
  +\frac{1}{(2l+1)(2l-1)}
\frac{p_<^l}{p_>^{l-1}} \left(1  
 - \frac{2l-1}{2l+3} \frac{p_<^2}{p_>^2} \right)
  + O\left(Rp_>^2\right)  \right\} \, .
\eea
The first term is a constant and cannot be separated
from an $S$-wave short-range interaction; it can therefore 
be absorbed in $V_S$, and we drop it not to clutter notation. 
In the limit $R \to 0$, one is then left with
\beq
V_{L\, l}^{(1)}(p', p) =  - \frac{\pi^2 g}{2m M^2}\frac{1}{(2l+1)(2l-1)}
\frac{p_<^l}{p_>^{l-1}} \left(1  
 - \frac{2l-1}{2l+3} \frac{p_<^2}{p_>^2} \right)
\, .
\label{eqn:nlo_ldi}
\eeq

We take $M$ to be the characteristic scale of the underlying theory, 
$M = M_{hi}$, and $|g| \sim \lambda$.
In this case, $V_L^{(1)}$ is a correction to $V_L^{(0)}$
at large distances $r\simge 1/M_{hi}$ or, equivalently, small momenta 
$p_>\simle M_{hi}$.
We want the short-range component $V_S$ to be such
that an expansion in $Q/M_{hi}$ holds for observables
obtained from the $T$ (or $K$) matrix.
Accordingly, we split $V_S$, $T$ and $K$ as in Eq. (\ref{eqn:VLsplit})
with the superscript $^{(0)}$ ($^{(1)}$) denoting LO (NLO).
RG invariance is exact only if all orders are considered.
Once a truncation to a finite order is made, 
Eq. (\ref{eqn:rge}) can only be satisfied up to
terms that vanish as $\Lambda\to \infty$ and can
be absorbed in higher-order counterterms.
In the rest of the paper we address the question of which
terms should be included in the short-range potential 
at each order to ensure a perturbative expansion of observables
consistent with RG invariance.

\section{\label{sec:LO} $-1/r^2$ as LO long-range potential}

We first tackle the problem in LO, that is, we take $V_L^{(1)}=0$.
The $S$-wave renormalization of an attractive inverse-square potential, 
$-1/r^2$, has been dealt with in coordinate- and momentum-space
in Refs. \cite{beane,coon} and \cite{hammer}, respectively. 
Here we present a new approach in momentum space, for any partial wave $l$.
Our $l=0$ results reproduce known results \cite{beane,coon,hammer}.

\subsection{Singularity of $-1/r^2$}
To see the origin of the peculiarities of a singular potential,
we start by taking $V_S^{(0)}=0$.
In this case we can write the integral Eq. (\ref{eqn:LSlnoccRED})
in a simplified form
\bea
t_l(px, p) &=& - \frac{\lambda}{2l+1}\, 
    \left\{x^{-(l+1)} 
    \left[\theta(x-1)\, 
          -\int_0^{x} dy \frac{y^{l+2}}{y^2 - 1- i\epsilon} t_l(py, p)
    \right]\right.\nonumber\\
&&  \left.\qquad\qquad
     +x^l \left[
            \theta(1-x)\, 
            -\int_{x}^{\Lambda/p} dy \frac{y^{1-l}}{y^2 - 1- i\epsilon} 
                    t_l(py, p) \right]\right\} \, .
\label{eqn:LSlnoccLLO}
\eea
In this form, we see that the only dimensionful parameter
is $\Lambda$, and scale invariance is evident
in the limit $\Lambda\to \infty$ (if it existed).

\begin{figure}
\includegraphics*[scale=0.7]{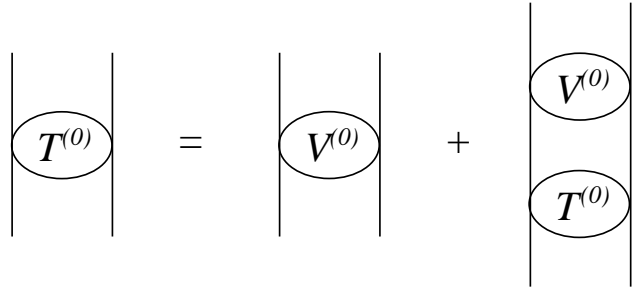}
\caption{\label{fig:t0} LO amplitude $T^{(0)}$ as an iteration of the LO
  potential $V^{(0)}$.}
\end{figure}

The validity of a perturbative expansion in powers
of $\lambda$ can be estimated
by comparing the first two terms,
\beq
-\frac{\lambda}{2l+1}
\eeq
and
\beq
- \frac{\lambda^2}{(2l+1)^2} \left( \frac{1}{2l+1} +
i\frac{\pi}{2} \right) \, ,
\eeq
in the expansion of the on-shell $t_l(p,p)$  
in Eq. (\ref{eqn:LSlnoccLLO}).
We see that perturbation theory in $\lambda$
works well when $l$ is high enough. 
Conversely, if 
\beq
l \simle l_p \equiv \frac{\lambda \pi}{4} - \frac{1}{2} \, ,
\label{eqn:lp}
\eeq
Eq. (\ref{eqn:LSlnocc}) has to be solved for
non-perturbatively. 
The LO amplitude $T^{(0)}$ is in this case obtained from
an exact solution of Eq. (\ref{eqn:LSlnocc}) with the LO potential,
as shown diagrammatically in Fig.~\ref{fig:t0}. 

In order to study the non-perturbative regime, we note that
the inverse-square potential,
Eq. (\ref{eqn:pot_mom}), has the interesting property
\beq
- {\bvec{\nabla}'}^2 V_L^{(0)}(\bvec{p'}, \bvec{p}) = -\frac{\lambda}{2m}
(2 \pi)^3 \delta^3(\bvec{p'}-\bvec{p}) \, ,
\eeq
where $\bvec{\nabla}' \equiv \frac{\partial}{\partial \bvec{p}'}$. Using this
property we can convert the Lippmann-Schwinger equation, Eq. (\ref{eqn:LS0}),
into a differential equation,
\beq
{\bvec{\nabla}'}^2 T(\bvec{p}', \bvec{p}) 
+ \frac{\lambda}{p'^2 - p^2  - i\epsilon} T(\bvec{p}', \bvec{p}) 
= \frac{4 \pi^3 \lambda}{m} \delta^3(\bvec{p}' - \bvec{p}) \, ,
\label{eqn:DEvec}
\eeq
or, decomposed onto partial waves,  
\beq
x\frac{\partial^2}{\partial {x}^2} \left(x t_l(px, p)\right) 
+ \left[\frac{\lambda x^2}{{x}^2 - 1 - i\epsilon}
- l(l+1) \right] t_l(px, p) 
= \lambda \, \delta(1-x) \, .
\label{eqn:DEl}
\eeq

Treating $p$ as a parameter, let us consider the solution of 
Eq. (\ref{eqn:DEl}). If $t_l(0, p)$ is assumed finite, which can be
inferred from the integral equation, then $t_l(px, p)$ for $x< 1$ is
determined up to a coefficient that is a function of $p$,
\beq
t_l(px, p) = \mathcal{N}_l(p/\Lambda) \, x^l \; 
{_2F_1} \left(
\frac{1}{2}\left(l+\frac{1}{2}-i\nu_l\right),
\frac{1}{2}\left(l+\frac{1}{2}+i\nu_l\right),
\frac{3}{2} + l;
x^2 \right) \, ,
\label{eqn:IRsol}
\eeq
where ${_2F_1}(\alpha, \beta, \gamma; z)$ is a hypergeometric function, and
\beq
\nu_l = \sqrt{\lambda - (l+1/2)^2} \, .
\label{eqn:def_nul}
\eeq
The pre-factor $\mathcal{N}_l$ must be a function
of $p/\Lambda$ on dimensional grounds. It is necessary for calculating the 
on-shell $t_l(p,p)$. 
When $x > 1$, the solution is the linear combination of two
generic solutions,
\bea
t_l(px, p) &=& A_l (p / \Lambda)  \, x^l \; 
{_2F_1} \left(
\frac{1}{2}\left(l+\frac{1}{2}-i\nu_l\right),
\frac{1}{2}\left(l+\frac{1}{2}+i\nu_l\right),
\frac{3}{2} + l; x^2 \right) \nonumber \\
&& + B_l (p / \Lambda) \, x^{-l-1} \; {_2F_1} \left(
-\frac{1}{2}\left(l+\frac{1}{2}+i\nu_l\right),
-\frac{1}{2}\left(l+\frac{1}{2}-i\nu_l\right),
\frac{1}{2} - l; x^2 \right)
\label{eqn:UVsol}
\eea
with coefficients $A_l$ and $B_l$ that can also be functions of $p/\Lambda$.
Since Eq. (\ref{eqn:DEl}) is inhomogeneous only at $x=1$,
there is no way to obtain $\mathcal{N}_l$ unless we determine the
ratio of $A_l$ to $B_l$ and then 
match to Eq. (\ref{eqn:IRsol}). 
This matching brings cutoff dependence to $\mathcal{N}_l$.
If the $x> 1$ solution were cutoff independent (up
to $O(p^2/\Lambda^2)$ corrections), one could expect
$\mathcal{N}_l$ and thus $t_l(p, p)$ to be constant, independent of $p$
---a consequence of scale invariance.

The $x> 1$ solution simplifies in the asymptotic region.
For $x\gg 1$, Eq. (\ref{eqn:DEl}) becomes
\beq
x\frac{\partial^2}{\partial {x}^2} (x t_l(px,p))
+\left[\lambda - l(l+1)\right] t_l(px,p)= 
O\left(\frac{t_l}{{x}^2} \right) \, .
\label{eqn:DElUV}
\eeq
This equation has two generic solutions,
\beq
t_l(px, p) \propto x^{\, \left(-\frac{1}{2} \pm i\nu_l\right)}
\left[1 +O\left(\frac{1}{{x}^2}\right) \right] \, . 
\label{eqn:UVsol0}
\eeq
If the potential is repulsive, {\it i.e.} $\lambda < 0$,
$\nu_l$ is imaginary; one of the two
solutions in Eq. (\ref{eqn:UVsol0}) has a positive power of
$x$, makes the second integral in Eq. (\ref{eqn:LSlnoccLLO}) divergent
when $\Lambda\to \infty$,
and has to be discarded. 
When the potential is attractive but not very strong,
namely $0 < \lambda < (l+1/2)^2$, $\nu_l$ is still imaginary;
neither solution produces a divergence in Eq. (\ref{eqn:LSlnoccLLO}) 
but the one with the bigger power  
must dominate over the other at high $x$. 
Hence in these two
cases we are left with only one solution for $x \gg 1$, up to a
coefficient dependent on $p$. 
With the ultraviolet behavior 
of the off-shell $t_l(p', p)$ decided one can in principle match to
Eq. (\ref{eqn:IRsol}), determining thus the on-shell amplitude
$t_l(p, p)$.

However, if the potential is attractive and sufficiently strong
to overcome the centrifugal barrier, that is, 
$\lambda \ge (l+1/2)^2$, then $\nu_l$ is real. 
For both solutions 
the second integral in Eq. (\ref{eqn:LSlnoccLLO}) is finite, but
oscillates with $\Lambda$. 
The two solutions have the
same magnitude but different phases.
In fact, the asymptotic expansion of
Eq. (\ref{eqn:UVsol}) gives
\beq
t_l(px, p) = \mathcal{N}'_l(p/\Lambda)  \, x^{-\frac{1}{2}}
\left[  \cos\left(\nu_l \ln \frac{xp}{\Lambda} + \theta_l \right)
+ \frac{1}{x^2} \cos\left(\nu_l \ln \frac{xp}{\Lambda} + \theta_l +
\beta_l \right) 
+ O\left(\frac{1}{{x}^4}\right) \right] \, ,
\label{eqn:asympcomb}
\eeq
where $\theta_l$ is an $l$-dependent constant and $\beta_l = \arg(-1
+ i\nu_l)$. 
Here we inserted the ultraviolet cutoff in the $\cos$ because 
the asymptotic dependence should be on $p'/\Lambda$.
(We absorbed a factor of $(p/\Lambda)^{-1/2}$ in $\mathcal{N}'$.)
The long-range potential, by itself, does not determine
the phase $\theta_l + \nu_l \ln p/\Lambda$ of the solution.
In fact, as we change the arbitrary cutoff $\Lambda$, the phase changes.
Equation (\ref{eqn:asympcomb}) is the asymptotic form
of the solution for $p'>p$, which is matched to Eq. (\ref{eqn:IRsol})
to determine $t_l(p, p)$. Thus, as $\Lambda$ changes, so
does the on-shell $t_l(p, p)$ and the observables obtained from it.

It is worth noting that the singularity of $-1/r^2$ depends on
the angular momentum $l$. 
Equation (\ref{eqn:def_nul}) implies that for any given $\lambda$
there is a critical $l_c$
above which $\nu_l$ is no longer real. Therefore the singularity
exists for
\beq
l< l_c \equiv \sqrt{\lambda} - 1/2 \, .
\label{eqn:lc} 
\eeq
In each of these waves, the attractive singular potential overcomes 
the centrifugal barrier, and observables are dependent on the arbitrary cutoff.

\begin{figure}
\includegraphics[scale=0.4]{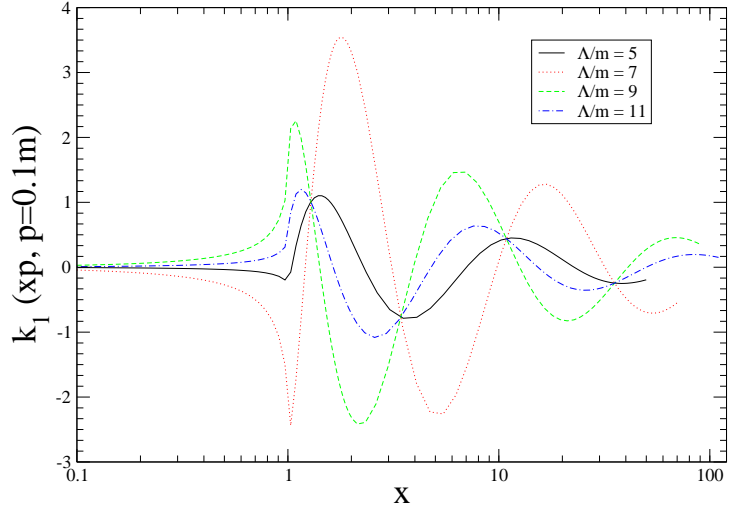}
\caption{\label{fig:koffnoct} $P$-wave reduced $K$ matrix
$k_1(xp,p)$ as function of $x$ for $p/m=0.1$
and various values of $\Lambda/m$,
when $V_{S\, 1}^{(0)} = 0$.
We have taken $\lambda =4.25$.}
\end{figure}

We can illustrate these facts with explicit numerical calculations.
Since the $S$ wave has already been studied in detail elsewhere \cite{hammer}, 
we focus on the $P$ wave, $l=1$.
We choose
$\lambda = 4.25$, which is strong enough
for the singularity to be present in the $P$ wave,
that is, $1<l_c<l_p$.
Figure \ref{fig:koffnoct} shows the numerical solution $k_1(px,p)$
of Eq. (\ref{eqn:LS-K}) for $p/m=0.1$,
at various values of the cutoff in units of the reduced mass,
$\Lambda/m$.
We see at $x\gg 1$ the asymptotic oscillations
of Eq. (\ref{eqn:asympcomb}): both the phase and
the amplitude depend on $\Lambda$.
The $\Lambda$ dependence 
propagates to smaller $x$
and results in
very different on-shell values $k_1(p, p)$.

Since the
choice of $\Lambda$ is arbitrary,
the LO on-shell $k_l^{(0)}(p, p)$ should be independent of
$\Lambda$ when $\Lambda$ is sufficiently large. But as shown  $k_l(p, p)$ with 
$V_S^{(0)}=0$
oscillates with
varying $\Lambda$ and the limit $\Lambda\to \infty$ is not well defined.
Therefore this $k_l(p, p)$ 
and the associated observables are meaningless.
RG invariance is not being respected in channels
where the attractive singular potential is treated non-perturbatively
and no counterterms are provided.

\subsection{\label{renor} Renormalization of $-1/r^2$}

The dependence on $\Lambda$ indicates that the $T$ matrix
is sensitive to short-range physics through virtual
states, that is, the loops resummed in the Lippmann-Schwinger equation,
Fig.~\ref{fig:t0}.
However, the high-momentum part of the loops cannot be distinguished
from contact interactions. The cutoff simply represents an
arbitrary division of where short-range physics is placed.
A model-independent accounting of this physics requires
the introduction of contact
interactions with parameters that depend on $\Lambda$ in such a way
that observables are independent of $\Lambda$.
Thus, with the technique of
renormalization one can get a cutoff-independent amplitude.
One has to renormalize each singular partial wave separately.
We shall show here that one counterterm in each singular partial wave
$l < l_c$ is necessary and sufficient to
renormalize the corresponding partial-wave amplitude. 
This extends the results of Refs. \cite{beane,coon,hammer} to waves beyond $S$.

For the $l$ wave, the contact interaction with fewest
derivatives in coordinate space can be written in a partial-wave projection
as
\beq
V_{S\, l}^{(0)}(p', p) = 
\frac{\pi^2}{m}\frac{C_l^{(0)}(\Lambda)}{(2l+1)}{p'}^l p^l \, ,
\label{eqn:ccl}
\eeq
where $C_l^{(0)}$ is a parameter.
Equation (\ref{eqn:LSlnoccRED}) becomes
\bea
t_l^{(0)}(px, p) &=& -\frac{1}{2l+1}
               \left\{ 
         \lambda x^{-(l+1)} \, \left[\theta(x-1) - \int_0^{x} dy \,  
        \frac{y^{l+2}}{y^2 - 1- i\epsilon} t_l^{(0)}(py, p)\right]
\right.
\nonumber \\
& & \left. \qquad\qquad + \lambda x^{l} \, \left[\theta(1-x) 
 - \int_{x}^{\Lambda/p} dy \, 
   \frac{y^{1-l} }{y^2 - 1- i\epsilon} t_l^{(0)}(py, p)\right]\right.
\nonumber \\
& & \left. \qquad\qquad +  C_l^{(0)} p^{2l+1} x^l
    \left[1 - 
    \int^{\Lambda/p}_0 dy \, 
    \frac{y^{l+2}}{y^2 - 1 - i\epsilon} t_l^{(0)}(py, p)\right]
    \right\} \, ,
 \label{eqn:LSlcc}
\eea
instead of Eq. (\ref{eqn:LSlnoccLLO}).

Renormalizing the $l$-wave amplitude means that the 
$\Lambda$ dependence of $C_l^{(0)}$ is such as to make
the on-shell
amplitude $t_l^{(0)}(p,p; \Lambda, C_l^{(0)}(\Lambda))$ independent of 
$\Lambda$ in
the large-$\Lambda$ limit:
\beq
\Lambda \frac{d}{d \Lambda} t_l^{(0)}\left(p, p; \Lambda, C_l^{(0)}(\Lambda) 
\right)=
O\left(\frac{p^2}{\Lambda^2}t_l^{(0)}\right) \; .
\label{eqn:rge0}
\eeq
For $-1/r^2$, we will show that the
half-off-shell $t_l^{(0)}(px, p; \Lambda, C_l(\Lambda))$ 
is also RG invariant in
the large-$\Lambda$ limit,
\beq
\Lambda \frac{d}{d \Lambda} t_l^{(0)}\left(px, p; \Lambda, C_l^{(0)}(\Lambda) 
\right)=
O\left(\frac{p^2}{\Lambda^2}t_l^{(0)}\right) 
\; .
\label{eqn:rgeoff}
\eeq
To justify this claim we take it as an {\it ansatz}, evaluate the RG
variation of the off-shell $t^{(0)}_l(px, p)$, and see if we can make it RG
invariant by controlling $C_l^{(0)}$ with varying $\Lambda$. To this end, we
take the total derivative with respect to $\Lambda$ of both sides of
Eq. (\ref{eqn:LSlcc}), assuming that 
$t^{(0)}(px, p; \Lambda, C_l^{(0)}(\Lambda))$ 
is RG invariant:
\beq
\frac{1}{2l+1}\left(\frac{xp}{\Lambda}\right)^l\left\{
    \left(\lambda -C_l^{(0)}\Lambda^{2l+1}\right) 
                        t_l^{(0)}(\Lambda, p)
   + (p\Lambda)^{l+1} \frac{d C_l^{(0)}}{d \Lambda}
  \left[
  1- I_{0,l} (p, \Lambda) \right] \right\} 
 =  O\left(\frac{p^2}{\Lambda^2}t_l^{(0)}\right) \; ,
\label{eqn:rhs0}
\eeq
where we defined 
\beq
I_{n,l}(p, \Lambda)=\int^{\Lambda/p}_0 dy \, 
  \frac{y^{2+2n+l}}{y^2 - 1 - i\epsilon} \, t_l^{(0)}(py,p)\; .
\label{eqn:Inl}
\eeq
Setting $x = \Lambda/p$ in Eq. (\ref{eqn:LSlcc}),
\beq
t_l^{(0)}(\Lambda, p) = -\frac{1}{2l + 1}
     \left(\lambda - C_l^{(0)} \Lambda^{2l+1}\right) 
     \left(\frac{p}{\Lambda}\right)^{l+1}
      \left[ 1- I_{0,l} (p, \Lambda) 
   \right] 
\; .
\label{eqn:LSlbc}
\eeq
Eliminating the expression in the square brackets in Eqs. (\ref{eqn:rhs0}) and
(\ref{eqn:LSlbc}),
\beq
\left(\frac{xp}{\Lambda}\right)^l
  \frac{t_l^{(0)}(\Lambda, p)}{\lambda-C_l^{(0)}\Lambda^{2l+1}} 
  \left[\frac{1}{2l+1} \left(\lambda -  C_l^{(0)} \Lambda^{2l+1}\right)^2
  -\Lambda^{2(l+1)} \frac{dC_l^{(0)}}{d \Lambda}  \right]
=  O\left(\frac{p^2}{\Lambda^2}t^{(0)}\right) \; .
\label{rhs1}
\eeq

Now, the left-hand side of Eq. (\ref{rhs1}) can be made of
$O(p^2 t^{(0)}/\Lambda^2)$ for all $x$ by properly choosing 
$C_l^{(0)}(\Lambda)$,
which is indeed consistent with the above {\it ansatz} that 
the half-off-shell
$t_l^{(0)}(px, p)$ is RG invariant, up to $O(p^2/\Lambda^2)$. 
This will be so if $C_l^{(0)}(\Lambda)$ satisfies the RG 
equation
\beq
\Lambda \frac{d C_l^{(0)}}{d \Lambda} \equiv \beta\left(C_l^{(0)}\right)
= \frac{1}{(2l+1) \Lambda^{2l+1}} 
        \left(\lambda- \Lambda^{2l+1} C_l^{(0)}\right)^2\, .
\label{eqn:rgecl}
\eeq
It is straightforward to solve for $C_l^{(0)}(\Lambda)$,
up to a boundary condition:
\beq
C_l^{(0)}(\Lambda) = - \frac{\lambda}{\Lambda^{2l+1}} 
  \frac{2l+1 - 2 \nu_l \tan [\nu_l\ln (\Lambda / \Lambda_{*l})]}
       {2l+1 + 2 \nu_l \tan [\nu_l\ln (\Lambda / \Lambda_{*l})]} \; ,
\label{eqn:rgc0}
\eeq
where $ \Lambda_{*l}$ is a dimensionful parameter
defined in such a way that the reduced coupling 
$\Lambda_{*l}^{2l+1}C_l^{(0)}(\Lambda_{*l})=-\lambda$.
It is determined by fitting to the measured $l$ partial-wave
amplitude. The log-periodic behavior of $C_l^{(0)}$ is the so-called limit
cycle~\cite{3bozos,beane,coon,hammer,limits}. 

Even with a non-zero $V_{S\, l}^{(0)}(p', p)$ the integral
equation (\ref{eqn:LSlnoccRED}) can be converted into Eq. (\ref{eqn:DEl}). 
The same argument goes through with $\Lambda_{*l}$ substituted
for $\Lambda$, dependence on which was eliminated, so that
the RG-invariant half-off-shell $t^{(0)}_l(p', p)$ 
in the $p' \gg p$ limit is
given by
\beq
t_l^{(0)}(px, p) = \mathcal{N}'_l(p/\Lambda_{*l})  \,
  x^{-\frac{1}{2}}  
             \, \left[  \cos\left(\nu_l \ln
	       \frac{xp}{\Lambda_{*l}}\right)
+ \frac{1}{x^2} \cos\left(\nu_l \ln \frac{xp}{\Lambda_{*l}} + \beta_l
\right) + O\left(\frac{1}{{x}^4}\right) \right] \, ,
\label{eqn:re_off_T0}
\eeq
instead of Eq. (\ref{eqn:asympcomb}). 
The phase is now fixed by $\Lambda_{*l}$, and observables are (nearly)
cutoff independent.

\begin{figure}
\includegraphics*[scale=0.4]{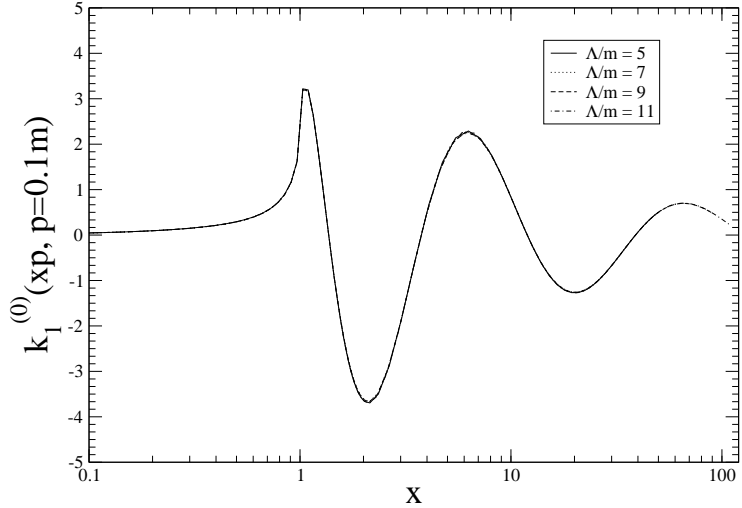}%
\caption{\label{fig:koffwithct} 
LO $P$-wave reduced $K$ matrix
$k_1^{(0)}(xp,p)$ as function of $x$ for $p/m=0.1$,
at various values of the cutoff $\Lambda$ in
units of the reduced mass $m$, with
$V_{S\, 1}^{(0)}$ introduced to absorb the cutoff dependence.
We have taken $\lambda =4.25$ and $\Lambda_{*1}/m = 0.2$.}
\end{figure}

To illustrate this we return to the $P$-wave example considered at the end of 
the previous subsection.
To remove the cutoff dependence observed in Fig.~\ref{fig:koffnoct},
we solve for the $K$ matrix again but now with an additional short-range
interaction $V_{S\, 1}^{(0)} = \pi^2 C_1^{(0)} p'p/3m$ put in and
allowed to change with $\Lambda$ in such a way that 
the on-shell $k_1^{(0)}(p, p)$ is independent of $\Lambda$.
Figure \ref{fig:koffwithct} shows that not only the on-shell
$k_1^{(0)}(p, p)$ but also the half-off-shell 
$k_1^{(0)}(p', p)$ are now independent of $\Lambda$,
in agreement with the previous argument.
In this calculation we chose $\Lambda_{*1}/m = 0.2$, 
but the results are qualitatively the
same for any $\Lambda_{*1}$.

The net effect of renormalization
is to replace $\Lambda$ by the physical quantities $\Lambda_{*l}$,
which parametrize short-range interactions
and are directly related to data.
The appearance of dimensionful parameters (through renormalization
with dimensionless parameters $\Lambda^{2l+1} C_l^{(0)}$ and $\lambda$) 
is an example of
dimensional transmutation.
These parameters introduce mass scales in the problem
and break the scale invariance of the long-range potential 
---an example of an anomaly.
(For an extensive discussion of this anomaly, see Ref. \cite{carlosanom}.)
Because of the log-periodic behavior, however,
a discrete scale invariance remains,
which has striking implications to observables \cite{hammer}.

If only one channel ($l=0$) is singular, the trade off
between $\Lambda$ and $\Lambda_{*0}$ is one-to-one;
using $\Lambda_{*0}$ is formally equivalent to treating 
$\Lambda$ as a fit parameter.
However, 
if more than one channel is singular at LO,
more than one short-range parameter is present in the solution.
Ignoring the counterterms and just fitting $\Lambda$ 
enforces a link between short-range parameters,
which might or might not be correct.
Regardless of whether this assumption is correct for a given
problem, it is nothing but a model for the short-range physics,
for it is a dynamical assumption that goes beyond
the symmetry content of the theory.
A model-independent treatment of short-range
interactions requires at LO one counterterm per singular channel
where perturbation theory does not apply.

\section{\label{sec:NLO} $\pm 1/r^4$ as NLO long-distance interaction}

\begin{figure}
\includegraphics*[scale=0.7]{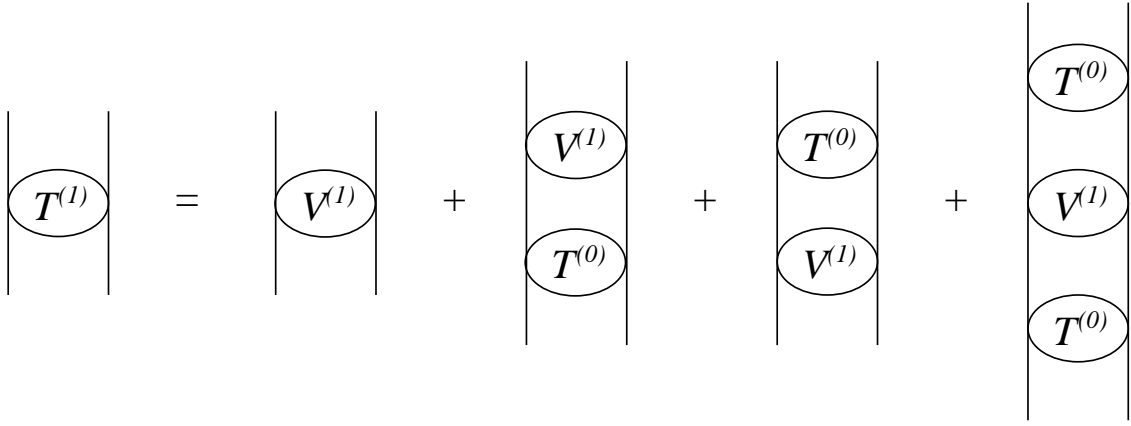}
\caption{\label{fig:t1} NLO amplitude $T^{(1)}$ in first-order
  distorted-wave Born approximation in the NLO potential $V^{(1)}$.}
\end{figure}

We now turn to the effects of singular perturbations 
on the LO singular potential and its counterterms.
We assume that the NLO long-range potential is given
by Eq. (\ref{eqn:nlo_ldi}), and ask which additional counterterms,
if any, have to be
supplied at NLO to make the result RG invariant.
With our choice of parameters, the long-range  potential,
Eq. (\ref{eqn:nlo_ldi}), is a correction to the LO long-range potential,
Eq. (\ref{Vl}), so the full NLO should have a small effect 
on observables.
Accordingly, we treat the NLO potential $V^{(1)}$ in perturbation theory,
that is, in first-order distorted-wave Born approximation.
The NLO amplitude $T^{(1)}$ has one insertion of the NLO potential,
see  Fig.~\ref{fig:t1}, that is,
\bea
t_l^{(1)}(p', p) &=& \frac{mp}{\pi^2} V_{l}^{(1)}(p', p) 
    - 2 \int_0^{\Lambda} dq \frac{q}{q^2 - p^2 - i\epsilon}
    \frac{mq}{\pi^2} V_{l}^{(1)}(p', q) t_l^{(0)}(q, p) \nonumber \\
    &&+ \int_0^{\Lambda} dq' \int_0^{\Lambda} dq t_l^{(0)}(p', q') 
        \frac{q'}{q'^2- p^2 - i\epsilon}
        \frac{m q'}{\pi^2} V_{l}^{(1)}(q', q) 
        \frac{q}{{q}^2 - p^2 - i\epsilon}
        t_l^{(0)}(q, p)
\; ,
\label{eqn:t01}
\eea
where we have used the symmetry of $V_{l}$ under exchange of its arguments.
Two insertions of $V^{(1)}$ (second-order perturbation theory) 
come at next order, where further contributions to
the long- and short-range potentials might exist.

For simplicity, we focus on the $S$ wave, $l=0$, in the 
following. 
Generalization to higher waves and higher orders is straightforward.

\subsection{Additional singularity}
In principle new cutoff dependence, possibly divergent in the 
$\Lambda\to \infty$ limit, will arise from the loops
in Fig.~\ref{fig:t1}.
Omitting for the moment any additional contact interactions in NLO, 
that is, taking $V_S^{(1)}=0$, let us 
consider 
the $\Lambda$ dependence of the on-shell amplitude. 
Taking the total derivative of Eq. (\ref{eqn:t01})
with respect to $\Lambda$ and using
the fact that the off-shell $t_0^{(0)}(px, p)$ is RG invariant
up to suppressed terms (see Eq. (\ref{eqn:rgeoff})), we find 
\bea
\Lambda \frac{d}{d \Lambda}  t_0(p, p) 
&=& - \frac{g \Lambda^2}{M^2}  \left(1- \frac{p^2}{\Lambda^2}\right )^{-1} 
    \left\{1 - I_{0,0}(p,\Lambda)
    + \frac{p^2}{3\Lambda^2} \left[1 - I_{1,0}(p,\Lambda)\right]\right\}
     t_0^{(0)}(\Lambda, p) \nonumber\\
&&+ O\left(\frac{p^4}{M^2\Lambda^2} t_0^{(0)}\right) \, ,
\label{eqn:woctms}
\eea
where the integrals $I_{n,l}(p,\Lambda)$ were defined in Eq. (\ref{eqn:Inl}).
Using Eqs. (\ref{eqn:LSlbc}) and (\ref{eqn:re_off_T0}) we can write
\beq
[1 - I_{0,0}(p, \Lambda)]\, t_0^{(0)}(\Lambda, p) =
{\mathcal{N}}_0^{'2} (p/\Lambda_{*0}) 
\left[ 
  G_0\left(\frac{\Lambda}{\Lambda_{*0}}\right) 
+ \frac{p^2}{\Lambda^2} G_1\left(\frac{\Lambda}{\Lambda_{*0}}\right) 
+ O\left(\frac{p^4}{\Lambda^4} 
\right) \right] \; ,
\eeq
where $G_0$ and $G_1$ are two $p$-independent,
oscillating functions of $\Lambda/\Lambda_{*0}$.
A similar form can be obtained for the other term,
\beq
\frac{p^2}{\Lambda^2}[1 - I_{1,0}(p, \Lambda)]t_0^{(0)}(\Lambda, p) =
{\mathcal{N}}_0^{'2} (p/\Lambda_{*0}) 
\left[ 
  H_0\left(\frac{\Lambda}{\Lambda_{*0}}\right) 
+ \frac{p^2}{\Lambda^2} H_1\left(\frac{\Lambda}{\Lambda_{*0}}\right) 
+ O\left(\frac{p^4}{\Lambda^4}\right) 
\right] \; ,
\eeq
in terms of two other oscillating 
functions $H_0$ and $H_1$ that do not depend on $p$.
This can be seen from the cutoff dependence of
$I_{1,0}(p, \Lambda)$ 
in the large-$\Lambda$ limit.
We can write
\beq
I_{1,0}(p,\Lambda)= I_{0,0}(p,\Lambda)
+ \int_{\Lambda'/p}^{\Lambda / p} dy \, y^2 \, t_0^{(0)}(py, p) 
+ \int_0^{\Lambda'/p} dy \, y^2 \, t_0^{(0)}(py, p) \; ,
\eeq
where $\Lambda'$ is a scale above which the asymptotic expansion of
$t_0^{(0)}$, Eq. (\ref{eqn:re_off_T0}), is valid. 
Since the $\Lambda$ dependence of the second integral is at most
$\propto p^2/\Lambda^{2}$ (see Eq. (\ref{eqn:rgeoff})),
$H_0$ and $H_1$ are given by the first integral.

Thus,
Eq. (\ref{eqn:woctms}) can be written as 
\bea
\Lambda \frac{d}{d \Lambda}  t_0(p, p) 
&=& - \frac{g}{M^2} {\mathcal{N}}_0^{'2}(p/\Lambda_{*0}) 
\left\{\Lambda^2 \left[
    G_0\left(\frac{\Lambda}{\Lambda_{*0}}\right) 
   + \frac{1}{3}H_0\left(\frac{\Lambda}{\Lambda_{*0}}\right) \right]
\right. \nonumber \\
&& +\left. p^2
\left[G_1\left(\frac{\Lambda}{\Lambda_{*0}}\right) 
    + G_0\left(\frac{\Lambda}{\Lambda_{*0}}\right) 
    +\frac{1}{3}H_1\left(\frac{\Lambda}{\Lambda_{*0}}\right) 
    +\frac{1}{3}H_0\left(\frac{\Lambda}{\Lambda_{*0}}\right) \right] 
+ O\left(\frac{p^4}{\Lambda^2}\right) 
\right\}.
\label{eqn:woctms1}
\eea
We find that there are two types of $\Lambda$-dependent terms 
(modulated by ${\mathcal{N}}_0^{'2}$) that 
do not vanish in 
the large-$\Lambda$ limit:
an energy-independent term whose
oscillatory behavior gets enhanced by an arbitrary factor $\Lambda^2$,
and a term that introduces further cutoff dependence proportional
to the energy. 
This stronger cutoff dependence 
is just the 
momentum-space reflection of the higher singularity of $V_L^{(1)}$.
Results become sensitive to the physics at the smaller distances 
where $V_L^{(1)}$ overcomes $V_L^{(0)}$. 
To account for this physics in a model-independent way, new counterterms
are needed.

\subsection{NLO renormalization}

Since $V_L^{(1)}$ is more singular than $V_L^{(0)}$ by two powers
of momenta (c.f. Eqs. (\ref{Vl}) and (\ref{eqn:nlo_ldi})),
we expect, on the basis of dimensional analysis, that new counterterms
with up to two extra derivatives or powers of $\Lambda$ will be required.  
Indeed, the two types of $\Lambda$ dependence in Eq. (\ref{eqn:woctms1}) 
suggest that we need two new
counterterms, one being possibly just a correction to $C_0^{(0)}$.
In that case, the running of $C_0$ is changed at NLO. 
For clarity, we split $C_0$ into two pieces,
$C_0= C_0^{(0)} + C_0^{(1)}$, where $C_0^{(0)}(\Lambda)$ 
has the LO running given in Eq. (\ref{eqn:rgc0})
and $C_0^{(1)}(\Lambda)$ has an NLO running to be determined.
However, this counterterm cannot be expected to
eliminate the energy-dependent term.
That requires a new counterterm $D_0^{(1)}$, which
represents the leading energy dependence 
of the short-range physics, and whose running $D_0^{(1)}(\Lambda)$
should also be determined
from the requirement of RG invariance of the $T$ matrix.
These arguments suggest that the NLO short-range potential is
\beq
V_{S\, 0}^{(1)}(p',p) = \frac{\pi^2}{m} 
                  \left[ C_0^{(1)} + D_0^{(1)}(p^2 + {p'}^2) \right] 
\, .
\label{eqn:nlocct}
\eeq

Including these terms
and using Eq. (\ref{eqn:LSlbc}), we get, instead of Eqs. (\ref{eqn:woctms})
and (\ref{eqn:woctms1}),
\bea
\Lambda \frac{d}{d \Lambda}  t_0^{(1)}(p, p) 
&=& -\left\{ \Lambda^2 \left[1 - I_{0,0}(p,\Lambda)\right] 
    \left[\left(\frac{g}{M^2}+\frac{2C_0^{(1)}}{\Lambda}+2D_0^{(1)}\Lambda 
          \right) \frac{\Lambda^2}{\Lambda^2 - p^2} 
     -\frac{1}{C_0^{(0)}\Lambda - \lambda}
     \frac{d C_0^{(1)}}{d \Lambda} \right] \right. \nonumber \\
&& \left. + \;  p^2 \left[1 - I_{1,0}(p,\Lambda)\right]
     \left[\left(\frac{g}{3 M^2} + 2D_0^{(1)}\Lambda \right)
       \frac{\Lambda^2}{\Lambda^2 - p^2}
     - \frac{2\Lambda^2}{C_0^{(0)} \Lambda - \lambda} 
     \frac{d D_0^{(1)}}{d\Lambda} \right] \right\} t_0^{(0)}(\Lambda, p)
\nonumber \\
&& + O\left(\frac{p^4}{M^2 \Lambda^2} t_0^{(0)}\right) \, ,
\nonumber \\
&=& -{\mathcal{N}}_0^{'2}(p/\Lambda_{*0}) 
     \frac{\Lambda^2 }{M^2}
     \left[R(\Lambda) + \frac{p^2}{\Lambda^2}  S(\Lambda) 
     + O\left(\frac{p^4}{\Lambda^4}\right)  \right]\; ,
\label{eqn:woctmsfull}
\eea
where
\bea
\frac{1}{M^2} R(\Lambda) &=& G_0\left(\frac{\Lambda}{\Lambda_{*0}}\right) 
             \left(\frac{g }{M^2} +\frac{2C_0^{(1)}}{\Lambda}  
             +2D_0^{(1)}\Lambda - \frac{1}{C_0^{(0)}\Lambda - \lambda}
             \frac{d C_0^{(1)}}{d \Lambda}\right)\nonumber
     \\
&& + \; H_0\left(\frac{\Lambda}{\Lambda_{*0}}\right) 
     \left(\frac{g}{3 M^2} + 2D_0^{(1)}\Lambda  
           - \frac{2\Lambda^2}{C_0^{(0)} \Lambda - \lambda} 
             \frac{d D_0^{(1)}}{d\Lambda} \right) 
\eea
and
\bea
\frac{1}{M^2} S(\Lambda) &=& G_1\left(\frac{\Lambda}{\Lambda_{*0}}\right) 
             \left(\frac{g }{M^2} +\frac{2C_0^{(1)}}{\Lambda}  
             +2D_0^{(1)}\Lambda - \frac{1}{C_0^{(0)}\Lambda - \lambda}
             \frac{d C_0^{(1)}}{d \Lambda}\right)\nonumber
     \\
&& + \; H_1\left(\frac{\Lambda}{\Lambda_{*0}}\right) 
     \left(\frac{g}{3 M^2} + 2D_0^{(1)}\Lambda  
           - \frac{2\Lambda^2}{C_0^{(0)} \Lambda - \lambda} 
             \frac{d D_0^{(1)}}{d\Lambda} \right) 
    \nonumber \\
&& + \; G_0\left(\frac{\Lambda}{\Lambda_{*0}}\right) 
    \left(\frac{g }{M^2}+\frac{2C_0^{(1)}}{\Lambda}+2D_0^{(1)}\Lambda \right)
   + \; H_0\left(\frac{\Lambda}{\Lambda_{*0}}\right) 
     \left(\frac{g}{3 M^2}+2D_0^{(1)}\Lambda \right) 
\, .
\eea
It is clear that  $C_0^{(1)}$ and $D_0^{(1)}$
generate terms with the same types of
$\Lambda$ dependence as in Eq. (\ref{eqn:woctms}). 
Therefore, by
controlling $C_0^{(1)}$ and $D_0^{(1)}$, we can make 
both $R(\Lambda)=0$ and $S(\Lambda)=0$, and the NLO amplitude 
$t_0^{(1)}(p,p)$ becomes RG invariant up to $O(p^4 / M^2 \Lambda^2)$.

Because 
the RG equations of the NLO counterterms 
$C_0^{(1)}(\Lambda)$ and $D_0^{(1)}(\Lambda)$ 
are not particularly illuminating,
we turn to numerical experiments.
We take $\lambda = 2$, $g = 1$  and $M = 0.5m$.
To test RG
invariance, we define the
fractional NLO correction 
\beq
X(p, \Lambda) = 
\left|\frac{k_0^{(1)}(p, p; \Lambda)}
           {k_0^{(0)}(p, p; \Lambda)}\right| \, . 
\eeq

\begin{figure}
\includegraphics*[scale=0.4]{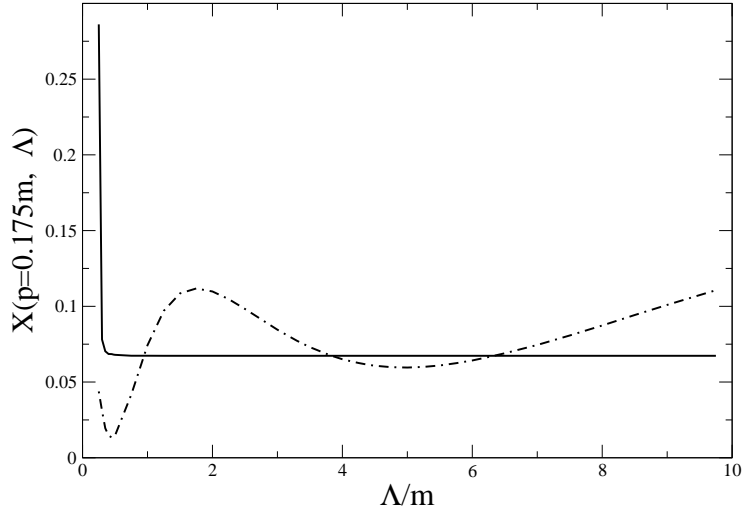}
\caption{\label{fig:X0175m} Fractional NLO correction $X(p, \Lambda)$
at $p=0.175m$ as function of $\Lambda/m$. 
The dot-dashed line is made by setting $D_0^{(1)}=0$ while
the solid line employs both $C_0^{(1)}$ and $D_0^{(1)}$.
We have taken $\lambda =2$, $g = 1$, and $M/m = 0.5$.}
\end{figure}

We first show that, in agreement with the previous argument,
renormalization cannot be done with $C_0^{(1)}$ alone.
We take as ``datum'' $k_0(0.1m, 0.1m) = -1.05$. 
In LO, $C_0^{(0)}(\Lambda)$ is determined so as to reproduce
this datum. The energy dependence is a prediction of the
theory.
In NLO, the additional terms in the potential will
make the theory deviate from the datum, unless
$C_0^{(1)}(\Lambda)$ is fitted to preserve agreement with the datum.
We thus solve the Lippmann-Schwinger equation
with various cutoffs, and
find
$C_0^{(1)}(\Lambda)$ such as to yield the datum.
In Fig. \ref{fig:X0175m} 
the dot-dashed line shows the fractional
NLO correction $X(p, \Lambda)$ as function of $\Lambda$ for
$p=0.175m$. 
It oscillates 
as $\Lambda$ increases and does not show sign of convergence.
One concludes that $C_0^{(1)}$
by itself does not renormalize the NLO amplitude.

We now repeat the calculation but including both
$C_0^{(1)}$ and $D_0^{(1)}$. Since two parameters are involved in the
fit, $k_0(0.1m, 0.1m) = -1.05$ and $k_0(0.15m, 0.15m) = -0.34$ 
are used as ``data''. 
The second data point is chosen as a 5\% displacement of
the value of $k_0^{(0)}(0.15m, 0.15m)$, to ensure that at low momenta
NLO represents a small effect on observables.
The cutoff dependence of the fractional NLO correction at
$p=0.175m$ is shown as the solid line in Fig. \ref{fig:X0175m}.
The plateau in the solid line
supports the hypothesis that the 
counterterms in Eq. (\ref{eqn:nlocct})
indeed renormalize the NLO amplitude.

\begin{figure}
\includegraphics*[scale=0.4]{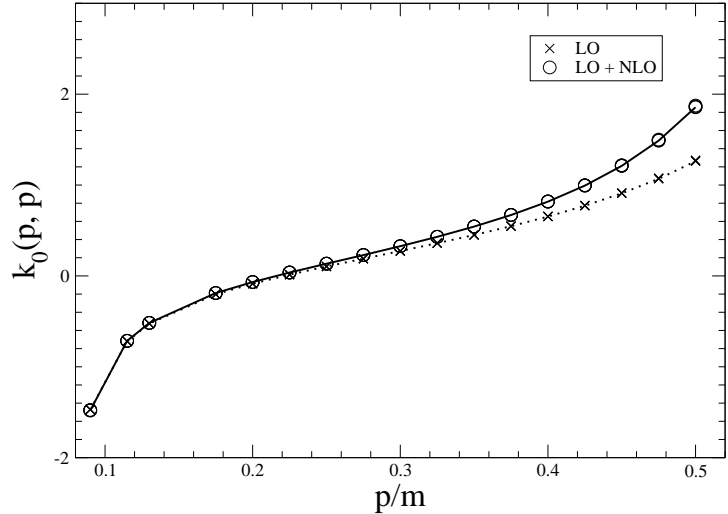}
\caption{\label{fig:loplusnlo} 
Reduced on-shell $K$ matrix $k_0(p, p)$ in the $S$ wave 
as a function of $p/m$,
with $\Lambda/m = 5.5, 6.5, 7.5$ and $8.5$:
leading order (crosses) and up to next-to-leading order (circles).
We have taken the same parameters as for the solid line in
Fig. \ref{fig:X0175m}.}
\end{figure}

With the data chosen above we compute the energy dependence of the
amplitude. 
Figure \ref{fig:loplusnlo} shows the energy dependence of both LO and 
LO+NLO reduced $K$ matrices.  Both sets of results are computed
with four different cutoffs $\Lambda/m = 5.5, 6.5, 7.5$ and $8.5$:
there are four data points on each spot in the figure.
The fact 
that the points for different cutoffs are indistinguishable
indicates that the amplitudes are being properly renormalized.
It is seen that the NLO correction is small for $p \ll M$
and fails for $p \sim M$, as it should. 

\section{\label{sec:discussion} Discussion}

In the previous sections we tackled a number of issues in the
rich physics of singular potentials using a simple toy model
$-1/r^2\pm 1/r^4$. 
Most of our results are independent of this particular choice,
but some are a consequence of the classical scale invariance
of the LO potential.
We discuss here some of the limitations and generalizations
of these results to an LO potential of the form
$-\lambda/2m M_{lo}^{n-2} r^n$, $n\ge 2$ and $\lambda=O(1)>0$,
perturbed by a more singular potential. 
Here $M_{lo}$ is a characteristic scale that sets the curvature
of the long-range potential. Clearly, at distances $r\sim 1/M_{lo}$
there can be a balance between kinetic repulsion and long-range
attraction, so this is a natural size for a bound state
or resonance to have. Accordingly,
a momentum $Q\sim M_{lo}$ is the most interesting resolution to consider.

In Sect. \ref{sec:LO} we presented a new method for the study of
the Lippmann-Schwinger equation with a singular potential,
in which we transformed it to a differential equation,
Eq. (\ref{eqn:DEvec}).
We would like to point out that this method 
can be applied to all $1/r^{2m}$ ($m$ an integer) potentials,
in which case we obtain a generalization 
of Eq. (\ref{eqn:DEvec}) that involves
the $m$th-order Laplacian  
  \footnote{The resultant differential equation resembles
  a Schr\"odinger equation, since the
  operator $r^2$ is just the Laplacian in the momentum representation. 
  It is not clear to us if a similar transformation exists
  for any type of long-range potential.}. 
The corresponding solutions then have 
similar properties.
An extension of the calculations of this paper to the more
general case is under investigation \cite{lbw}.

We have shown that angular momentum plays
an important, double role through the repulsive effects of the 
centrifugal barrier.
The emergence of the remarkable phenomena we discussed comes from
the competition between the singular potential and the centrifugal barrier.
When $n=2$, for a given singular potential strength, there is
a critical angular momentum $l_c$, Eq. (\ref{eqn:lc}), above which
the effective radial potential is no longer attractive,
and the particles are prevented from probing short-range physics.
In these waves, the problem is well defined in LO without a
short-range counterterm. 
Conversely, in waves with $l<l_c$, a counterterm is required in 
every wave, in agreement with a general argument \cite{behncke68}.
Such a critical angular momentum is a particularity of $n=2$.
For $n>2$, the singular potential will overcome the centrifugal
barrier (with $l>0$) at some distance 
$r_l\sim (\lambda/l(l+1))^{1/(n-2)}/M_{lo}$.
Therefore, the two particles will want to get as close as
allowed by short-range physics. To make observables minimally
sensitive to this short-range physics, a single
counterterm is required in every wave. 
Only recently was this found \cite{nogga_timmermans_vankolck,spaniards1} 
in the more complicated case of the direction-dependent $1/r^3$
potential originating from one-pion exchange between nucleons.

Regardless of the existence of $l_c$, the centrifugal barrier weakens 
(for $n=2$) or dominates (for $n>2$)
over the singular potential at large distances. 
The larger the $l$, the weaker the effective radial potential 
in a partial wave. 
This leads for $n=2$ to a second value for $l$,
$l_p$ in Eq. (\ref{eqn:lp}), above which the potential can be treated
in perturbation theory. 
When $n>2$, there is always a spatial region $r< r_l$
where the singular potential is strong, but its size decreases with $l$.
Whether the potential can be treated in perturbation theory in a given
partial wave depends then crucially on the range of momenta $Q$
that we are interested in probing. For a given $Q$,
a sufficiently large $l$ exists where the support of the 
strong potential is effectively a short-range effect.
For short-range potentials, barring fine-tunings leading to
bound states or resonances at threshold, 
perturbation theory holds in higher waves \cite{goldberger}.
In the nuclear case below the QCD mass scale, 
it was found that $l_p\approx 3$ \cite{nogga_timmermans_vankolck}.

It is thus a general feature of singular potentials that 
a single counterterm, Eq. (\ref{eqn:ccl}),
is needed in all waves where the potential
is sufficiently attractive.
For $n=2$, the presence of dimensionful parameters $\Lambda_{*l}$
breaks scale invariance,
but discrete scale invariance remains in the form of limit cycle.
Our LO results in the $S$ wave are essentially the same as in
the three-body system with short-range interactions 
\cite{3bozos, coverass2}.
For $n>2$, the dependence of counterterms on the cutoff
is slightly more complicated, but still oscillatory \cite{beane,mary}.

The renormalization of $-1/r^n$ suggests a different power counting
than what one would expect from naive dimensional analysis (NDA) 
or naturalness.
Based on NDA one would expect
that $C_l$ in Eq. (\ref{eqn:ccl}) scales as
\beq
C_l \sim \frac{1}{M_{hi}^{2l+1}} \, . 
\label{eqn:naiveLO}
\eeq
This is reasonable if $\lambda$ is outside of the singular region or 
if the Born
approximation is valid.
But if this is not the case the necessity of
renormalization promotes $C_l$ to the same size as the
long-range potential. The renormalized $C_l$ can be thought of 
as scaling instead with the resolution $Q$ at which the potential
is considered valid,
\beq
C_l^{(0)} \sim \frac{\lambda}{Q^{2l+1}}\, .
\label{eqn:realLO}
\eeq
The corresponding short-range interaction, Eq. (\ref{eqn:ccl}),
then scales as $1/mQ$, which is the same scaling
as the long-range potential.

In Sect. \ref{sec:NLO} we investigated the effects of
a perturbation in the form of a singular potential with
two more powers of momenta than the LO potential:
a $(Q/M_{hi})^2$ perturbation.
In the nuclear case, it has been suggested \cite{nogga_timmermans_vankolck}
and disputed \cite{spaniards2} that the additional
counterterms are those indicated by NDA,
in this case, those with two more powers of momenta.
We found here that a perturbative treatment of the corrections 
is indeed consistent with this NDA expectation.

Therefore, once the failure of NDA is corrected at LO, Eq. (\ref{eqn:realLO}), 
power counting is formulated as usual.
In the particular example considered here, Eq. (\ref{eqn:nlocct}),
\beq
C_0^{(1)} \sim \frac{g Q}{M_{hi}^2} \, , \qquad 
D_0^{(1)} \sim \frac{g}{Q M_{hi}^2} 
\, . 
\label{eqn:realNLO}
\eeq
As a consequence, the NLO short-range interaction, Eq. (\ref{eqn:nlocct}),
scales as $g Q/m M_{hi}^2$, just as the NLO long-range potential.

We can also use the 
arguments of Sect. \ref{sec:NLO}
to determine the NLO counterterms 
if the NLO long-distance potential is
another potential than $1/r^4$. For example, in the case of
$1/m M_{hi}^4 r^6$ as NLO, whose Fourier transform is $\sim Q^3/m M_{hi}^4$, 
one expects to need 
\beq
C_0^{(1)} + D_0^{(1)}({p'}^2 + p^2) + E_0^{(1)} {p'}^2 p^2
\eeq
as NLO counterterms in the $S$ wave, based on the qualitative
arguments of Sect. \ref{sec:NLO}. In this case
\beq
C_0^{(1)} \sim \frac{g Q^3}{M_{hi}^4}\, , \qquad 
D_0^{(1)} \sim \frac{g Q}{M_{hi}^4}\, , \qquad 
E_0^{(1)} \sim \frac{g}{Q M_{hi}^4}\, , 
\eeq
so that the short-range interaction scales
as $g Q^3/m M_{hi}^4$. 

This state of affairs is perhaps not surprising. 
NDA was developed based on experience with perturbation theory.
It does fail for an attractive singular potential, but only when
the potential is treated non-perturbatively in LO.
Once this unforeseen event is incorporated in the power counting,
the perturbative treatment of the corrections conforms to NDA,
relative to the corrected LO.

\section{\label{sec:summary} Summary and Outlook}

We have used $-1/r^2\textrm{(LO)} \pm 1/r^4\textrm{(NLO)}$ as an example to
demonstrate how to build effective theories based upon singular
potentials. The key point is to understand the power counting of
contact interactions. The correct power counting
scheme should consist of the minimum set of contact interactions that
renormalizes scattering amplitudes including the long-range potentials. 
Due to this intrinsic
relation between renormalization and power counting it was found that
the sizes of contact interactions are different from what
is expected by naive dimensional analysis.

A new approach to the renormalization of a $-1/r^2$ potential was presented. 
It was shown
that the singularity of $-1/r^2$ depends on angular momentum. The
region of singularity entangles with that of validity of the Born
approximation. In each partial wave where the LO potential is resummed to all
orders, a single counterterm is needed for renormalization.
The NLO potential can be treated as a (distorted-wave)
perturbation, and the minimum set of NLO short-range counterterms that
are needed to renormalize the NLO amplitude can be determined by
estimating the superficial cutoff dependence with the asymptotic LO
$T$ matrix. Analytical arguments were supplemented by numerical
evidence. 

It is one of the main conclusions of this paper that the power counting, 
a key ingredient of any effective theory, cannot be decided 
solely on the basis of naive
dimensional analysis in the case of (non-perturbative) singular
potentials. One has to rely on explicit checks of RG invariance, either
analytical or numerical, in order to test any
proposed power counting scheme.
Yet, once LO has been understood, perturbative corrections
do not violate naive dimensional analysis with respect to LO.

Besides nuclear forces
\cite{weinberg,others,towards,nogga_timmermans_vankolck,spaniards1,germans,spaniards2,sameold}, 
there may be other applications of 
effective theories of singular potentials.
For example, the $1/r^2$ potential in two dimensions is relevant for the
interaction of a neutral atom with a charged wire 
(see, {\it e.g.}, Ref. \cite{denschlag}),
while it appears in three dimensions 
with an angle-dependent coefficient when
a charge is in the field of a polar molecule 
(see, {\it e.g.} Ref. \cite{leblond}). 
Our results could be readily applied to long-range corrections
in these systems
\footnote{It is well known that a three-body system with short-range 
 pairwise interactions can be mapped \cite{efimov} 
 ---when the two-body $S$-wave scattering length $a_2 \to \infty$--- into a
 two-dimensional $-1/r^2$ potential problem. 
 Our method can thus be adapted to this system as well.
 The counterterm necessary to renormalize the $-1/r^2$ problem
 in LO represents a three-body force in the original system,
 while our NLO analysis is related to the controversy \cite{nnlodispute} 
 regarding the renormalization of higher-order three-body forces in three-body
 systems when both $a_2$ and the two-body effective range $r_2$ are finite. 
 However, if $a_2$ is finite the mapping is complicated \cite{birse} 
 even in the case $r_2=0$. 
 An investigation of whether our method can be usefully applied
 to the three-body case of interest is worthwhile but beyond the scope of the
 present manuscript.}. 
With extensions \cite{lbw}, it could also be applied to
the (long-range)
electron-atom interaction, which is often divided into $-1/r^4$, $1/r^6$ and
higher terms \cite{frank_land_spector}. 
In all these cases, one can construct effective theories with a well
defined power counting that incorporate renormalization-group ideas.

\acknowledgments 

This work was supported in part by the U.S. Department of Energy (BL, UvK),
by a Galileo Circle Scholarship from the College of Science
of the University of Arizona (BL),
by the Nederlandse Organisatie voor Wetenschappelijk Onderzoek (UvK),
and 
by Brazil's FAPESP under a Visiting Professor grant (UvK).
UvK would like to thank the hospitality of 
the Kernfysisch Versneller Instituut at Rijksuniversiteit Groningen,
the Instituto de F\'\i sica Te\'orica of the
Universidade Estadual Paulista,
and the Instituto de F\'\i sica of the Universidade de S\~ao Paulo
where part of this work was completed.

\end{document}